\documentclass[preprint]{aastex}

\shorttitle{Cosmic rays removal}
\shortauthors{Pych}

\begin{document}

\title{A Fast Algorithm for Cosmic Rays Removal from Single Images}

\author{Wojtek~Pych}
\affil
{David Dunlap Observatory, University of Toronto \\
P.O.~Box 360, Richmond Hill, Ontario, Canada L4C 4Y6 \\
and \\
Copernicus Astronomical Center \\
Bartycka 18, 00--716 Warszawa, Poland}

\email{pych@camk.edu.pl}

\begin{abstract}

We present a method for detecting cosmic rays in single images.
The algorithm is based on simple analysis of the histogram of the image
data and does not use any modeling of the picture of the object.
It does not require a good signal to noise ratio in the image data.
Identification of multiple-pixel cosmic-ray hits is realized by running
the procedure for detection and replacement iteratively.
The tests performed by us, show that the method is very effective, when
applied to the images with the spectroscopic data. It is also very fast
in comparison with other single image algorithms found in astronomical
data processing packages.
Practical implementation and examples of application are presented. 

\end{abstract}


\section{INTRODUCTION}
\label{sec1}

Cosmic ray hits cause defects in all astronomical images obtained
with CCD detectors.
A relatively efficient approach to removing traces of cosmic rays
from such images is to use multiple frames of the same object and then
combine them with some algorithm for rejection of the outlying data.
Methods of this type have been
presented among others by \citet{Shaw1992} and \citet{Windhorst1994}.
They may be found in most astronomical data processing packages.

In many cases we are not able to obtain multiple images of the same
object or the required time resolution prevented us from using multiple
image methods. 
A straightforward method that could be used in such cases would be sigma
clipping. In practice it is almost impossible to find a good detection
threshold for such an algorithm. In fact, it may leave some obvious cosmic
rays untouched, while giving numerous spurious detections and rejecting
valid data at the same time.
Other types of single image algorithms may also be found in most
astronomical data processing packages.
Most of these methods rely on the sharpness of cosmic rays, relative to
the atmospheric smoothing of real images. These methods are
based on some interactive learning techniques
\citep{Murtagh1992,Salzberg1995} or involve a special fitting
of a model representing a real image of a star with a superimposed cosmic
ray \citep{Rhoads2000}. These methods were designed to correct direct
images of the stellar objects and we find them not suitable for
spectral images.

The exposure times of the spectra are usually longer than
direct images and range from 600~seconds to 1800~seconds.
Considerable amounts of cosmic ray events accumulate in the images
during such exposures.
We have constructed a simple and straightforward algorithm for
detecting cosmic rays in single images. Our method does not need any
model of the shape of the image features themselves.
We analyze the histograms of
pixel counts in small sub-frames in order to detect pixels deviating by
some factor from the bulk of the pixels under consideration.

We would like to note here, that
low value bad pixels tend to hide weak cosmic rays by artificially making
the local variance larger than would be caused only by usual sources of
noise in the image: readout noise, dark current and the Poisson
noise associated with background and object signal.
The cosmic rays have random distributions of their positions in the
images. On the other hand, bad pixels may be found as deviating pixels at
the same position in every frame taken with a given instrument.
Therefore we suggest to make bad pixel correction as a separate step of
the reduction process prior to cosmic rays correction.

We describe details of our algorithm in paragraph 2. In paragraph 3. we
describe an implementation and the results of our tests.
In paragraph 4. we present examples of application to real astronomical data.
Paragraph 5 contains the summary.

\section{The Algorithm}
\label{sec2}

The proposed algorithm is based on the idea that cosmic rays deposit a
portion of their energy in the pixels they hit, causing some extra
signal in these pixels. The signal coming from cosmic rays does not have
a Gaussian distribution.
This should reflect in the distribution of counts in the image affected
by cosmic rays.
The image as a whole may have a large range of signal levels in different
areas. We analyze relatively small sub-frames, to work with
a more concentrated local distribution of counts. In most cases this 
distribution should be rather compact. The cosmic rays appear as points
standing out in the high counts interval.

\begin{figure}
\figurenum{1}
\plotone{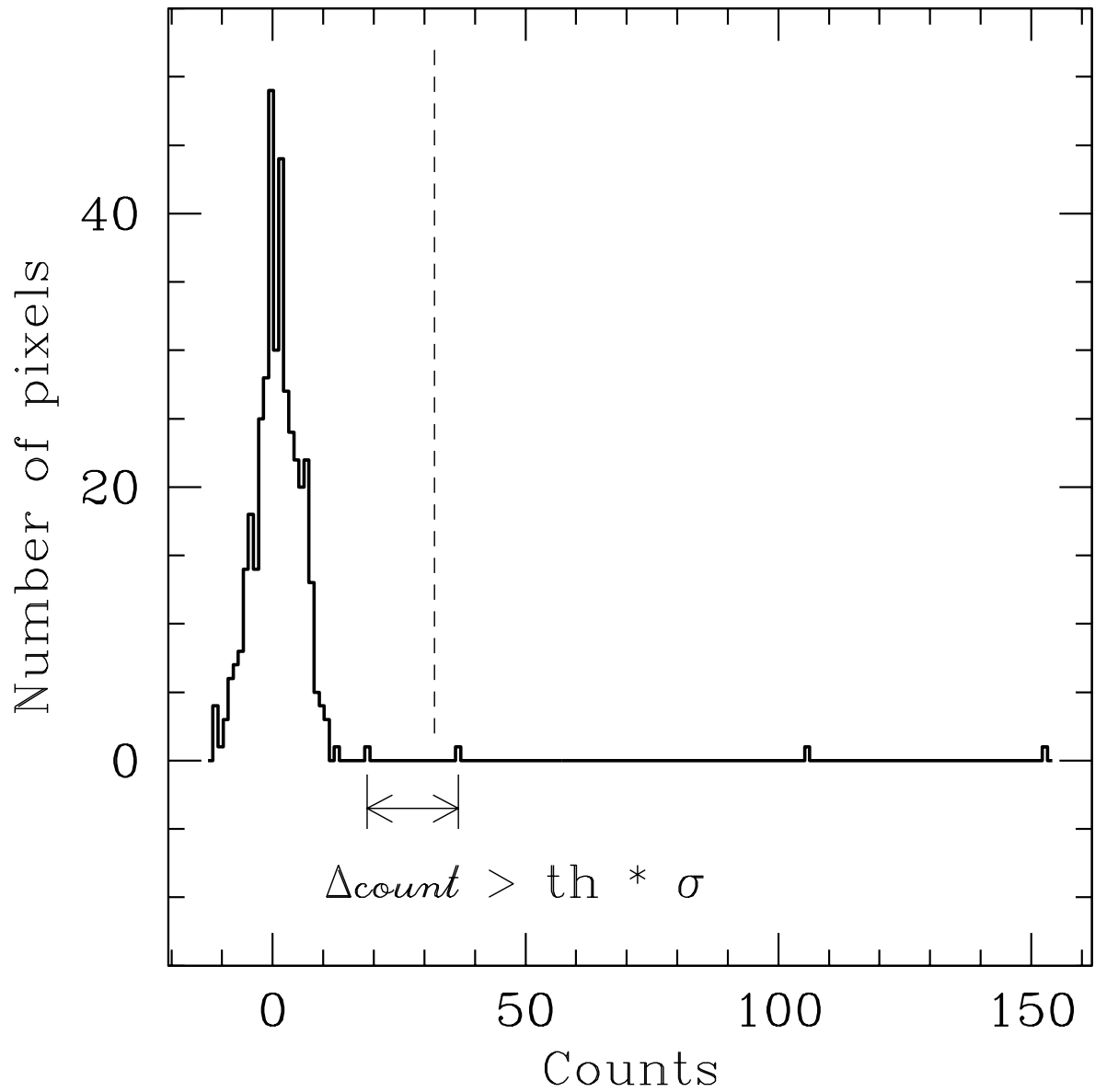}
\caption{Histogram used in a typical application of 
the cosmic-ray detection algorithm.
The vertical dashed line marks the lower limit of counts regarded as the
cosmic rays. The arrows mark the first gap in the histogram of the width
larger than the threshold.}
\label{fig1}
\end{figure}

The proposed algorithm of cosmic rays detection generally consists of
the following steps:
\begin{enumerate}
\item Select small size sub-frames, which cover the whole frame with
substantial overlapping. 
\end{enumerate}
In each sub-frame:
\begin{enumerate}
\setcounter{enumi}{1}
\item Calculate standard deviation of the  distribution of counts:
\begin{equation}
\sigma = \sqrt { {\sum{c_i^2} - {(\sum{c_i})^2 \over n}} \over {n} }
\end{equation}
\item Apply a single sigma clipping step to correct the estimate of
standard deviation for outlying pixels.
\item Construct a histogram of the distribution of counts.
\item Find the mode of the distribution of counts, i.e. the peak of the
  histogram.
\item In the interval of counts higher than the mode,
  find gaps in the histogram i.e. bins with zero data points.
\item Find the first gap which is wider than a threshold, which is the
  standard deviation multiplied by an arbitrary number (usually 3.0).
\item If such gap exists, flag pixels with counts lying above the gap as
  affected by cosmic rays.
\end{enumerate}

Figure \ref{fig1} illustrates an example of a histogram
with a few pixels identified as affected by cosmic ray events.

The next step consists of a replacement of the count numbers in the
flagged pixels. If we consider a single image, the information about the
real signal in the affected pixels is lost.
In many cases however, the characteristic scale of spacial
variations of signal are of the order of at least a few pixels. In such
cases, one may use some interpolation to replace missing pixel counts.
In our implementation, we decided to substitute the cosmic rays by the
average of the counts in the neighboring pixels.

Cosmic rays are often multiple-pixel events. High signal pixels may
screen neighboring pixels from detection. For this reason
we run the procedure of detection and cleaning cosmic rays iteratively.
The process is rapidly converging and usually there are no new detections
after 2 or 3 iterations. We have also introduced a parameter called
"growing radius", which tells the procedure to clean the pixels closer
than this radius to the detected one, even when they do not possess
enough signal to be flagged as affected by cosmic rays.

\section{Implementation and Test}
\label{sec3}

\begin{figure}
\figurenum{2}
\plotone{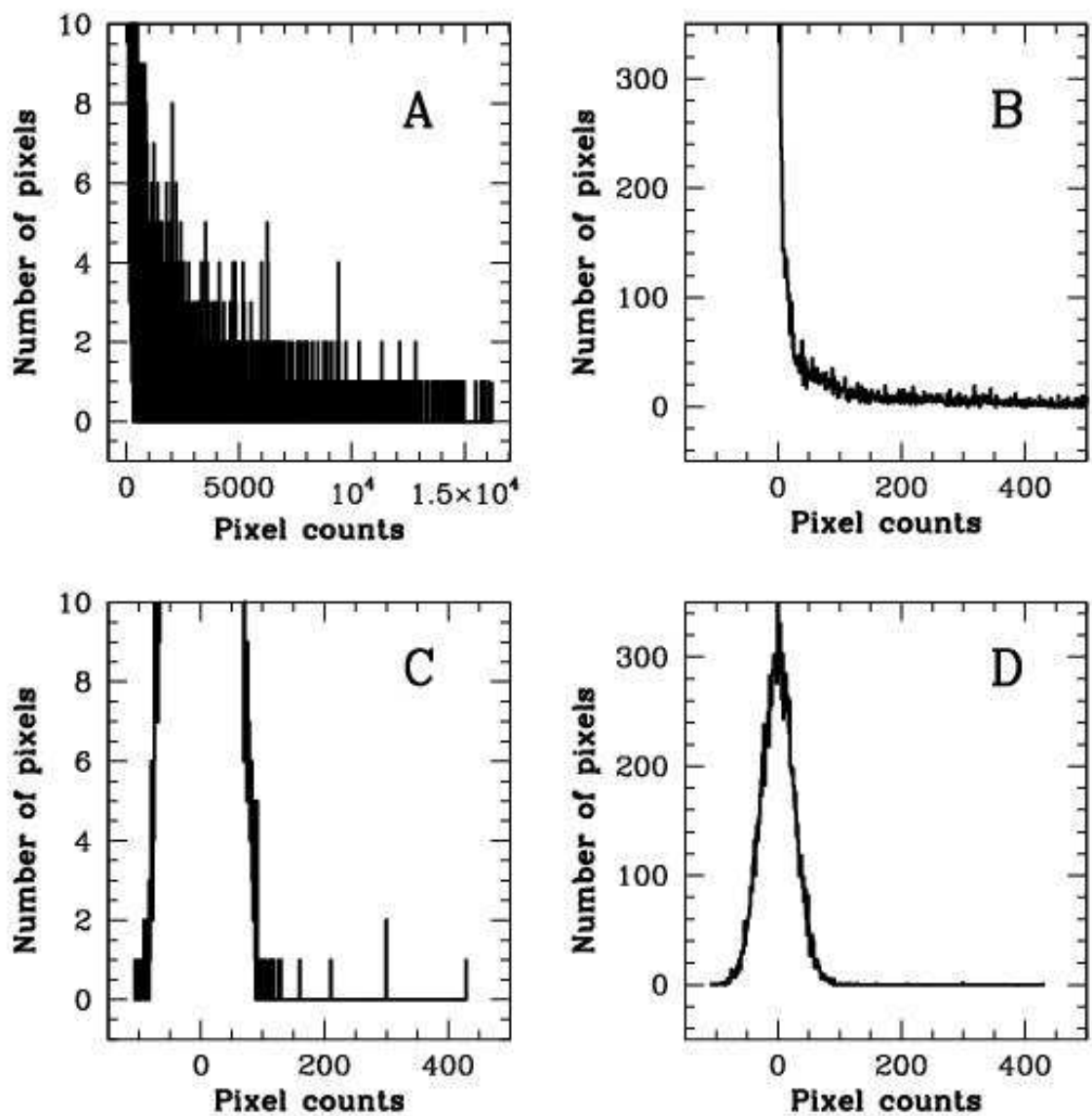}
\caption{Distribution of counts associated with cosmic rays. Panel A:
distribution of counts associated with high energy cosmic rays;
Panel B: distribution of counts associated with low energy cosmic rays;
Panel C: distribution of counts associated with undetected cosmic rays;
Panel D: distribution of residual counts after cosmic ray cleaning. See
text for details.}
\label{fig2}
\end{figure}

We have written a computer program, which implements the above described
algorithm\footnote{The C-source of our program may be downloaded
from the following Web-page: \url{http://www.camk.edu.pl/\~{ }pych/}}.
This program has several input parameters which allow the user to control
the process of detecting and cleaning cosmic rays.
These parameters define:
\begin{itemize}
\item
the size of the sub-frame box: X and Y dimensions of the rectangle,
\item
the threshold: the number by which the local variance is multiplied,
\item
the dispersion axis: the interpolation for bad pixel substitution is
calculated along this axis; if no axis selected, it is calculated
in the annulus around that pixel,
\item
the lower and upper radii of the annulus for the interpolation of data,
\item
the growing radius: the maximum distance of the pixels which are
to be corrected in the neighborhood of the flagged pixel.
\end{itemize}

Our program also creates a map of detected cosmic rays.
This map can be examined to check for any correlation with the real
data. The existence of such correlation indicates that the
detection threshold has been set too low and the real data has been
modified.

To check the capabilities of our method,
we have generated an artificial two-dimensional Echelle spectrum with the
size of \mbox{2048 x 4096 pixels}. For this purpose, we have used a procedure
from IRAF package: {\it noao.artdata.mkechelle}, with its default parameters.
We have also used the procedure {\it noao.artdata.mknoise}
to obtain two noise images. The first one consists of readout noise
(5 electrons) and a background of 500~counts with Poisson noise. The
second image consists of 1000 cosmic rays with maximum energy
30000~electrons.

Our program finished with zero detections on the noiseless image of the
Echelle spectrum. The result for the spectrum with the readout and Poisson
noise added was one detection over the whole frame.
Finally, we tested the procedure on a
spectrum with Poisson noise and cosmic rays added to the original
image of a spectrum. Our program has found all but one cosmic
ray event of peak greater than 200 counts.
The only event left, and with a peak of 430 counts, was located
in the vicinity of a trace of the spectrum and was concealed by this
signal.

Figure \ref{fig2} presents histograms of the counts distribution.
The upper-left panel (A) presents the original distribution of counts
from cosmic rays. The Y axis is cut at a value of 10 pixels to highlight
the distribution of high energy hits.
The upper-right panel (B) presents the distribution of
cosmic rays in the interval of small counts.
The lower panels present the distribution of counts in the image obtained
as a difference between the spectrum image after cosmic rays cleaning
and the image without cosmic rays added.
The Y-axis in the lower-left panel (C) is cut at a value of 10 pixels,
to show the distribution of pixels with undetected cosmic rays. The
features visible above 200 counts are the traces of a single undetected
cosmic ray event.
The lower-right panel (D) presents the distribution of the residual
counts. It has the shape of a Gaussian with a small 'high-end' tail,
produced by undetected cosmic rays.
The standard deviation of the residuals, $\sigma = 26.0$, is similar to
the variance of the original image without cosmic rays; $\sigma = 27.8$.
This similarity reflects the nature of the adopted substitution method
for the pixels affected by cosmic rays, which is simply a mean value
of the surrounding pixels.

\begin{figure}
\figurenum{3}
\plotone{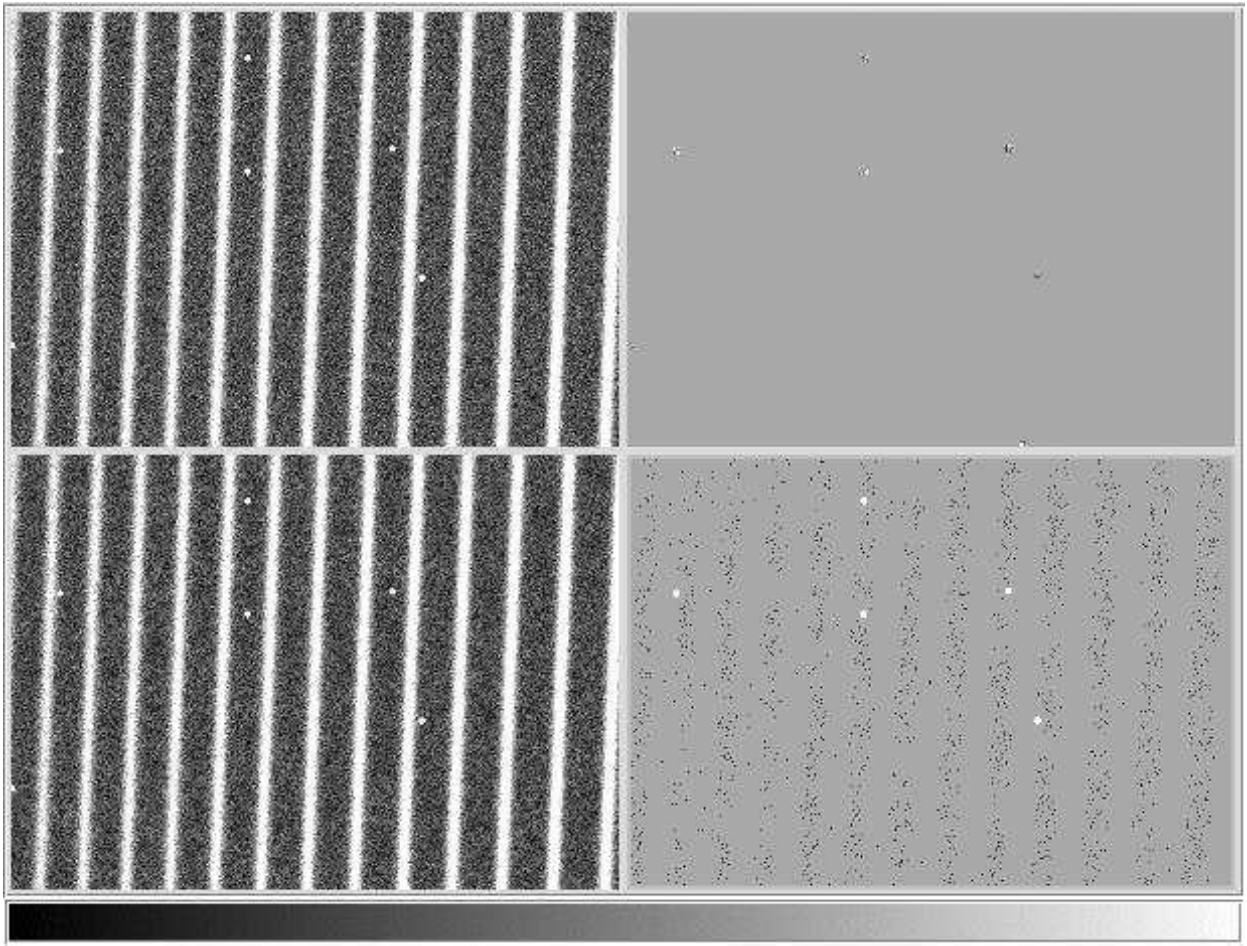}
\caption{Central fragments of our test images.
The upper-left panel
shows the artificial Echelle spectrum with the Poisson noise
and cosmic rays added.
The upper-right panel
shows the difference between the image cleaned by our program
and the original spectrum (without cosmic rays).
The lower-left panel
presents the same image as in the panel above, but after being
processed by the IRAF procedure {\it noao.imred.crutil.cosmicrays}.
The lower-right panel
shows the difference between the image obtained from 
the IRAF routine {\it noao.imred.crutil.cosmicrays}.
and the original spectrum (without cosmic rays).}
\label{fig3}
\end{figure}

Our tests were conducted on a PC with a 500~MHz Intel Celeron processor.
The program was compiled using a GNU C Compiler
(gcc version 3.2.2) under Red Hat Linux 9.0.
The CPU time needed to process a single frame: 2048~x~4096, 32~bit pixels,
was about 40~seconds. The CPU time required to process an image
depends linearly on the number of the image pixels.

We have also run the IRAF procedure {\it noao.imred.crutil.cosmicrays}.
The CPU time needed to process the same image was over 900~seconds.
In comparison with the above, our algorithm may be classified as a fast
one. The relative speed
results from the fact that the whole detection process does not require
extensive calculations.
Our test demonstrated that the IRAF routine was not able to remove
multiple pixel cosmic ray events. They were slightly modified, but
most of them remained in the frame.
At the same time, a large number of image pixels unaffected by cosmic
rays were modified.
We would like to stress here that our algorithm does remove cosmic rays
effectively while leaving almost all of the image data untouched.

Figure \ref{fig3} presents central fragments of our test images.
The upper-left panel
shows the artificial Echelle spectrum with the Poisson noise and cosmic
rays added.
The upper-right panel
presents the residual signal remaining after subtraction of the original
spectrum without cosmic rays from the image cleaned by our program.
The lower-left panel
presents the image after being processed by the IRAF procedure
{\it noao.imred.crutil.cosmicrays}. Note that the cosmic rays were not
removed.
The lower-right panel
shows the residual signal remaining after subtraction of the original
spectrum without cosmic rays from the image obtained
from the IRAF routine {\it noao.imred.crutil.cosmicrays}.
The scales of the images of the residual signals are the same.

\section{Examples of practical application}
\label{sec4}

\begin{figure}
\figurenum{4}
\epsscale{1.0}
\plotone{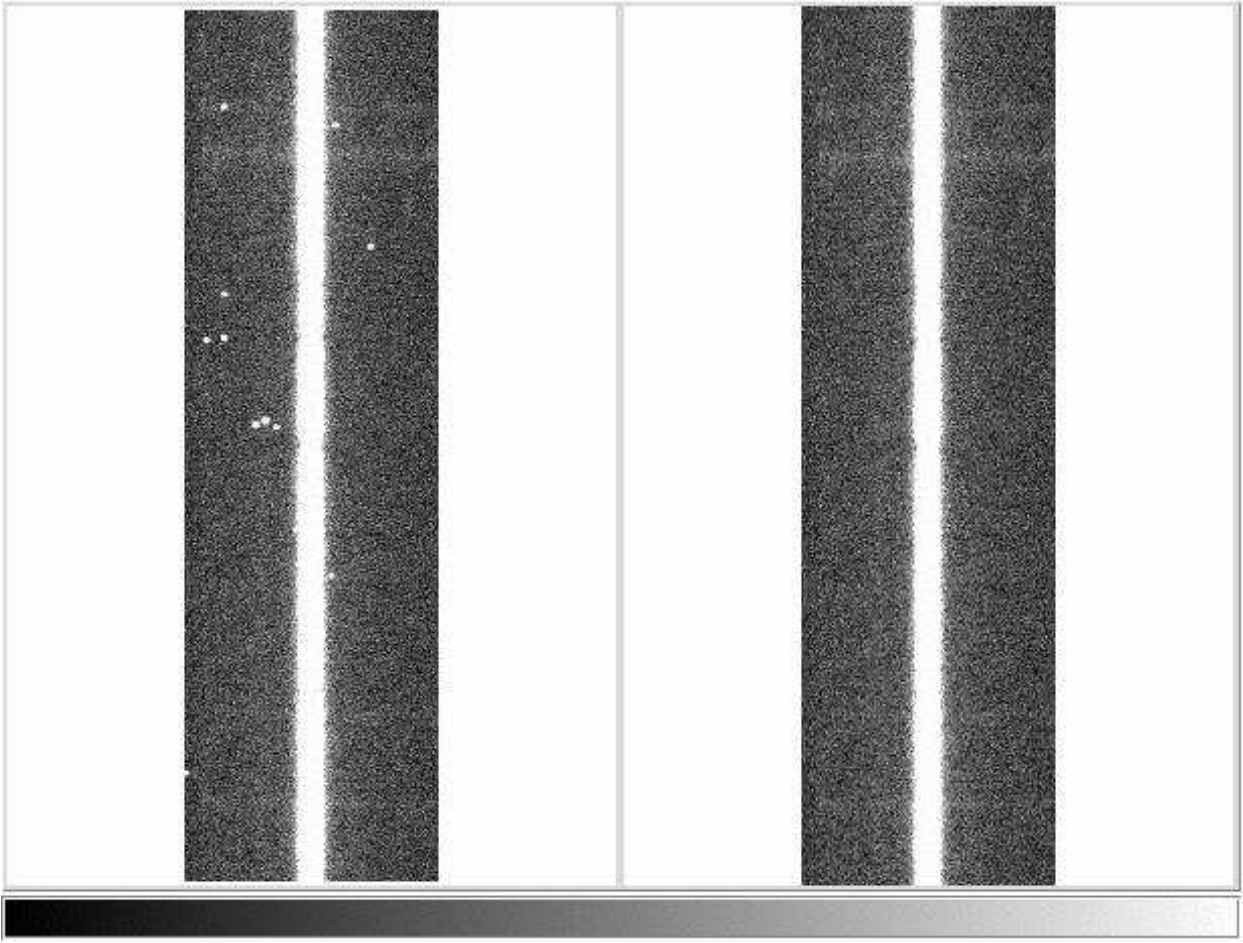}
\figcaption{Example of a frame with a long slit spectrum: 
before cosmic ray cleaning -- left panel;
after cosmic ray cleaning -- right panel.
The stellar signal is in the white band along each panel.}
\label{fig4}
\end{figure}

\begin{figure}
\figurenum{5}
\epsscale{1.0}
\plotone{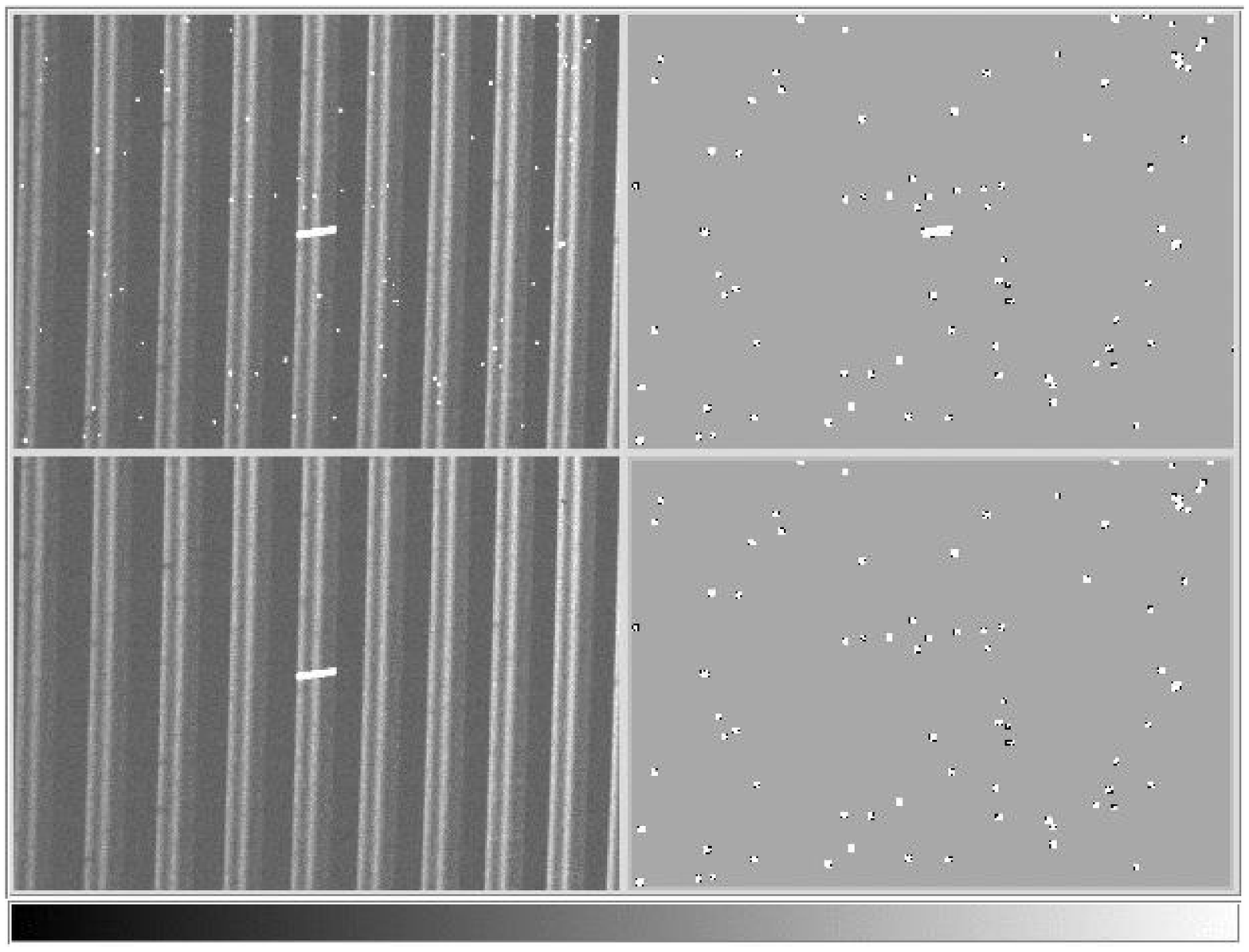}
\figcaption{Example of the application of our method on an Echelle
spectrum of a faint star.
The upper-left panel
shows a fragment of the spectrum of a 17~mag. star,
obtained in an 1800~second exposure with MIKE spectrograph
at the Las Campanas Observatory.
The upper-right panel
presents the map of cosmic rays detected in the image. A sky emission
line is visible in the center.
The lower-left panel
presents the same fragment of the spectrum as shown in the upper-left
panel, but after subtraction of the map of cosmic rays modified to
remove the sky emission line.
The lower-right panel
shows the map of cosmic rays with the sky emission line removed.}
\label{fig5}
\end{figure}

We have applied the above algorithm to several data sets.
Our general experience is that the method works very well on images of
both long-slit and Echelle spectra.
The algorithm was originally designed to work with the spectroscopic data.
Our method could be used to clean images of stellar
fields with a very wide Point Spread Function (PSF).
It works well in a sense that it does not produce too many
spurious detections. The limitation of the presented method for these
types of images comes from the fact that direct images usually have
large count variations within small scales. This produces a large
standard deviation of the counts and prevents cosmic rays in the
neighborhood of bright objects from being detected.
We suggest using the method proposed by \citet{Rhoads2000} in such
cases.

Figure \ref{fig4} illustrates an example result of cosmic ray removal in a
long-slit spectrum. A fragment of the
spectrum of a 9.2~mag. variable star, obtained in a 1200~second
exposure with the 1.88-m telescope
at the David Dunlap Observatory, University of Toronto, is shown.
The left panel
presents the original image (after bias subtraction and flat-field
correction), while the right panel shows the cleaned image. No evident
cosmic rays are identified by eye in the cleaned frame.

Our method works best on low signal images.
For the faint stars however, we encounter another problem.
The night-sky emission lines, which are sharp features perpendicular to
the dispersion axis, may sometimes be strong enough to be identified
as cosmic rays. Our solution to this problem is to edit the map of cosmic
rays, erase features identified as sky emission lines (replace the
counts with zeros) and then subtract a modified cosmic ray map from
the original image.

Figure \ref{fig5} illustrates an example of the application on an Echelle
spectrum of a faint star.
The upper-left panel
shows a fragment of the Echelle spectrum of a 17~mag. star,
obtained in an 1800~second exposure with MIKE spectrograph
(Magellan Inamori Kyocera Echelle) at the Las Campanas Observatory.
The upper-right panel
presents the map of cosmic rays detected in the image. A sky emission
line is visible in the center.
The lower-left panel
presents the same fragment of the spectrum as shown in the upper-left
panel, but after subtraction of the modified map of cosmic rays. The map
was edited to remove the sky emission line.
The lower-right panel
shows the map of cosmic rays with the sky emission line removed.

\section{Summary}
\label{sec5}

We have presented a cosmic-ray rejection algorithm based on a simple
analysis of the histogram of the image data.
The most important advantage of our
method is that it does not require modeling of the image data and
may {\it apriori} be applied to any type of well sampled image data.
We have checked that for the spectroscopic images, it is very effective
in detecting cosmic rays, while avoiding numerous spurious detections.
Our method does not require advanced and extensive computations, so it is
relatively fast.

The weak point of this approach, is that bright objects may shield cosmic
rays in their neighborhood from detection. This is caused by two factors.
First: the sensitivity to cosmic ray events is reduced at the locations of
bright objects because of the Poisson noise associated with the image
photons. Second: we look for bright spots and the legitimate object
may be brighter than a nearby comic ray trace.

Our suggestion is to use the algorithm presented above for spectroscopic
data whenever multiple image methods cannot be employed for cosmic rays
removal.

Future improvement to the presented method could be an introduction of
better interpolation for the replacement of pixels affected by cosmic
rays.

\acknowledgements

The author would like to thank Prof. S. Rucinski, J. R. Thomson and H.
DeBond for help in preparing this article.
This article has been prepared when W. Pych held the NATO Post-Doctoral
Fellowship administered by the Natural Sciences and Engineering Council
of Canada (NSERC).
The author acknowledges also the support from the Polish Grant KBN
2~P03D~029~23 and the NSERC research grant to S. Rucinski.

\end{document}